\documentstyle[12pt,aaspp]{article}  

\newcommand{\beq}{\begin{equation}}           % begin equation
\newcommand{\eeq}{\end{equation}}             % end equation

\begin{document}

\title{Probing Red Giant Atmospheres with Gravitational Microlensing}

\author{David Heyrovsk\'y, Dimitar Sasselov and Abraham Loeb}

\affil{Department of Astronomy, Harvard University\\ 60 Garden St., 
Cambridge, MA 02138, USA} 

\begin{abstract}

Gravitational microlensing provides a new technique for studying the
surfaces of distant stars. Microlensing events are detected in real time
and can be followed up with precision photometry and spectroscopy. This
method is particularly adequate for studying red giants in the Galactic
bulge. Recently we developed an efficient method capable of computing the
lensing effect for thousands of frequencies in a high-resolution stellar
spectrum. Here we demonstrate the effects of microlensing on synthesized
optical spectra of red giant model atmospheres. We show that different
properties of the stellar surface can be recovered from time-dependent
photometry and spectroscopy of a point-mass microlensing event with a
small impact parameter. In this study we concentrate on center-to-limb
variation of spectral features. Measuring such variations can reveal the
depth structure of the atmosphere of the source star.

\end{abstract}

\bigskip
\noindent
\bigskip
 
\centerline{Submitted to {\it The Astrophysical Journal}, February 1999}
 
\clearpage 
\section{Introduction}

Stars populating the red giant branch spend a substantial fraction ($\geq
5\%$) of their entire lifetime there. Red giants are thus a major
component of any galactic stellar population. In old systems, like
Galactic globular clusters, their contribution is even larger. The
parameters of the red giant branch have been traditionally derived from
multi-color photometry and interpreted using stellar atmosphere models
(e.g. Vandenberg \& Bell~1985; Kurucz~1992). These model atmospheres
involve one-dimensional integrations of semi-infinite slabs in hydrostatic
and radiative equilibrium. However, real atmospheres of red giants are
usually extended and the slab approximation is not adequate.  Despite
significant progress in computing extended spherically-symmetric models
(e.g., Scholz \& Tsuji~1984; Plez, Brett, \& Nordlund~1992; to name a
few), it is still difficult to use these improvements in massive grids of
models, which therefore continue to be computed as plane-parallel
(Houdashelt et~al.~2000). A number of other assumptions and
simplifications had to be usually made, some of which have not been
verified yet by direct observations. Examples include the treatment of
convective transport and small-scale velocity fields (e.g. micro- and
macro-turbulence). Similarly, stellar evolution models also depend on a
number of parameters which describe poorly understood physics, such as
convection near the surface of the star.

These problems have hindered significant improvement in the determination
of fundamental stellar properties over the past two decades. The angular
diameters of only a small number of stars have been measured up to now,
with an accuracy of $\sim$ 10\% at best (Armstrong et~al.~1995). Stellar
disk brightness distributions (limb darkening) have been inferred only for
a handful of stars at the level of a proof-of-concept (directly $-$
Mozurkewich et~al.~1991; or in binaries $-$ Andersen~1991). An exception
is provided by the direct $HST$ observation of the red supergiant $\alpha$
Ori, which revealed limb darkening as well as a bright region on the
surface (Uitenbroek, Dupree, \& Gilliland~1998). In general, the available
inferred results cannot be used to build a model atmosphere, as in the
case of the Sun. Due to this lack of an adequate observational basis, our
understanding of stellar light is implicitly tied to the solar atmosphere
model.
 
This situation is particularly unfortunate, since recently there has been
an increased demand for accurate stellar models in a number of
applications. For example, much improved color-temperature conversions are
needed for calibrating distance indicators and determining stellar ages.
Stellar population syntheses (e.g. for high-redshift galaxies) also rely
critically on the accuracy of current stellar models. Our Sun is not a
good standard for most of these applications.

The advent of large interferometric arrays (Keck interferometer, CHARA,
SIM, VLTI) will go a long way towards solving these problems, but these
complex and expensive facilities are still mostly in a preliminary
development phase. In the meantime, gravitational microlensing offers an
easily accessible, immediate, and inexpensive means for imaging at least
large stars, such as red giants. It also offers, by its nature, access to
stellar populations in the Galactic bulge and the Magellanic Clouds, which
are well beyond the reach of any interferometer.

By now there are more than 350 microlensing events detected towards the
Galactic bulge and the Large and Small Magellanic Clouds (see e.g. Alcock
et~al.~1997a, 1997b, 1997c; Becker et~al.~1998; Renault et~al.~1997;
Afonso et~al.~1999; Udalski et~al.~1997). During a Galactic gravitational
microlensing event the flux from a distant star is temporarily amplified
by the gravitational field of a massive dim object passing in the
foreground. The standard light curve in the point-source limit is
characterized by two observables: its peak amplitude and duration
(Paczy\'nski~1996). These observables physically depend only on the
lensing impact parameter and the Einstein radius crossing time of the
source. However, in events with a small impact parameter the light curve
is also affected by the resolved surface of the lensed star, thus
providing an excellent opportunity for stellar surface imaging
(Valls-Gabaud~1995, Sasselov~1996).

The first well documented case of resolved finite-size microlensing
effects is MACHO Alert 95-30 (hereafter M95-30), with the impressive
follow-up campaign by the GMAN and MACHO collaborations (Alcock
et~al.~1997d). The first spectroscopic observations of a binary lens event
in which the caustic crossed the face of the source star were reported by
Lennon et~al.~(1996). In a similar caustic-crossing binary event, Albrow
et~al.~(1999) recently determined limb darkening profiles of the lensed K
giant star for both observed spectral bands ($I$ and $V$).

Finite-size effects were first theoretically studied as methods to
partially remove the degeneracy of microlensing light curves. These
methods are based on alterations of the standard light curve (e.g.
Gould~1994, 1995; Nemiroff \& Wickramasinghe~1994; Witt \& Mao~1994;
Gould \& Welch~1996), resolved polarization (Simmons, Willis, \&
Newsam~1995; Simmons, Newsam, \& Willis~1995; Bogdanov, Cherepashchuk,
\& Sazhin~1996), spectral shifts due to stellar rotation (Maoz \&
Gould~1994), and narrow-band photometry of resonance lines (Loeb \&
Sasselov~1995). Here we want to put emphasis on the inverse problem $-$
using microlensing for studying stellar surface features and probing the
atmosphere of the source star.

The resolved stellar surface brightness distribution,
$B(\lambda,\vec{r}\,)$, can vary strongly with wavelength in selected
spectral regions - in the continuum as well as within spectral lines. The
time-dependent microlensing amplification then becomes
wavelength-dependent through $B(\lambda,\vec{r}\,)$. Photometry in
different bands and sets of spectra taken in the course of a microlensing
event can therefore be used to study the brightness distribution of the
source. Microlensing light curve chromaticity was first evaluated by
Witt~(1995), Valls-Gabaud~(1995) and Bogdanov \& Cherepashchuk~(1995).
Methods for obtaining the brightness distribution by light curve inversion
were studied by Bogdanov \& Cherepashchuk~(1996) and Hendry et~al.~(1998),
inversion error analysis was presented by Gaudi \& Gould~(1999). Few works
have been published so far on spectral effects in stellar microlensing.
These include studies of line profile changes in rotating giants (Maoz \&
Gould~1994; Gould~1997) and in circumstellar envelopes in bulk motion
(Ignace \& Hendry~1999). First results for selected spectral lines in a
simple model stellar atmosphere were presented by Valls-Gabaud~(1996,
1998).
 
In order to optimize observations of spectral effects in actual events,
the most practical approach is to theoretically predict the lensing effect
on the entire synthesized optical spectrum of the source in microlensing
alerts similar to M95-30. This way one can determine the most sensitive
spectral features before the event peaks. An efficient method and computer
code capable of dealing with the tens of thousands of frequencies involved
was developed by Heyrovsk\'y \& Loeb~(1997; hereafter Paper I). In this
work we use the method to explore point-mass microlensing effects on both
low and high resolution optical spectra, synthesized directly from
up-to-date red giant model atmosphere calculations.

The outline of this paper is as follows. In \S 2 we describe the model
atmospheres used in this paper. The effect of microlensing on the overall
spectrum of the source star as well as on individual spectral features is
presented in \S 3. Finally \S 4 summarizes the main conclusions and future
prospects of this work.

\section{Model Atmospheres of Red Giants}

In a stellar population with some spread of metallicities red giants
exhibit a range of temperatures. The cooler giants have atmospheres which,
unlike in the case of the Sun, are dominated by molecular compounds. Most
of the hydrogen gets locked into H$_2$ and most of the carbon into CO. As
the temperature drops, TiO, VO, and eventually H$_2$O become prominent.
This requires paying special attention to the equation of state and
background opacities (Hauschildt et~al.~1997; Kurucz~1992). Of great
importance is the treatment of convection, as transport by convection
affects a large part of the photosphere. Finally, for models with high
luminosity (and low surface gravity, log~$g$ $\leq$ 1), spherical geometry
must be adopted in the treatment of radiative transfer (Hauschildt
et~al.~1999). 

\begin{figure}[tb]
\plottwo{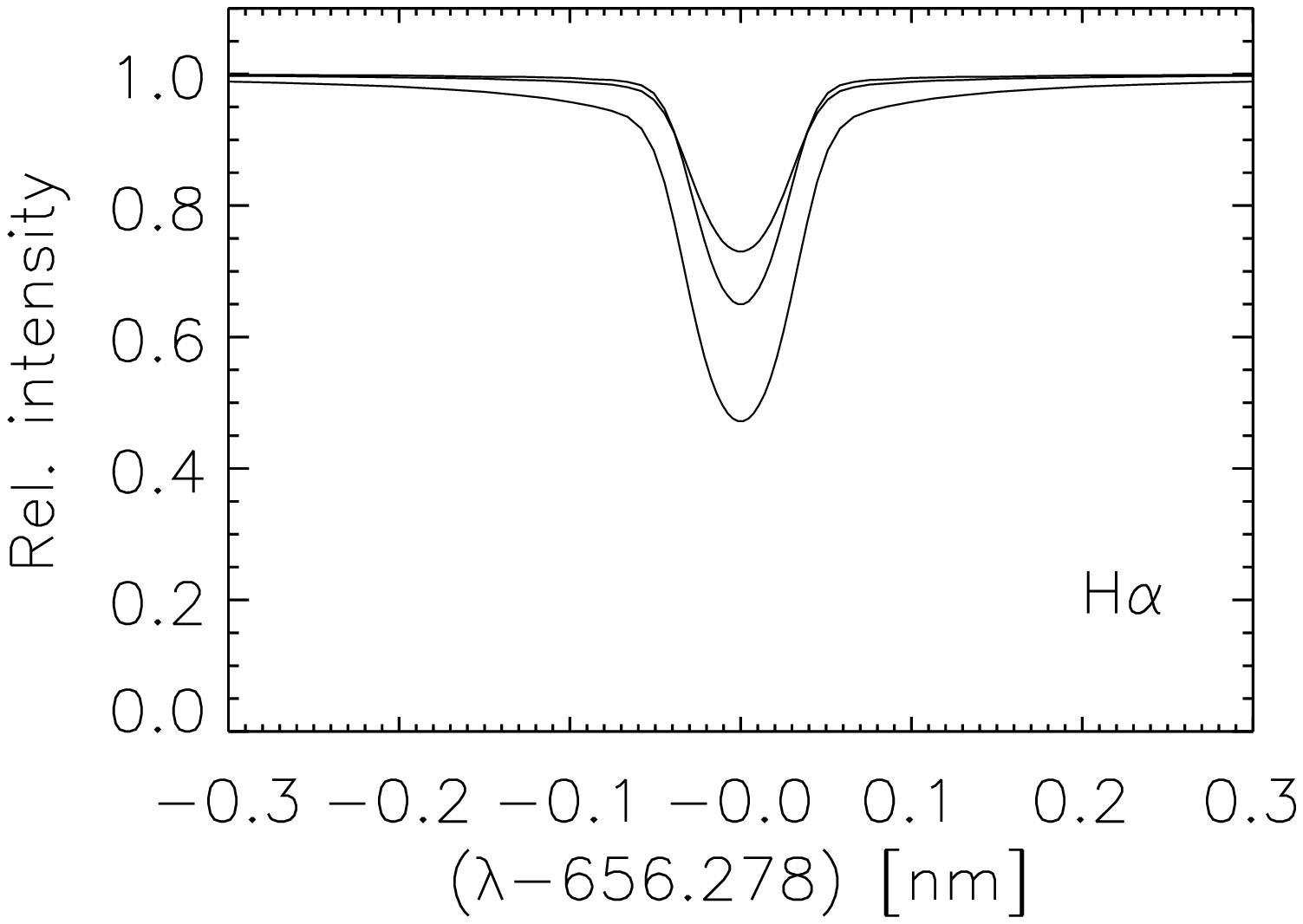}{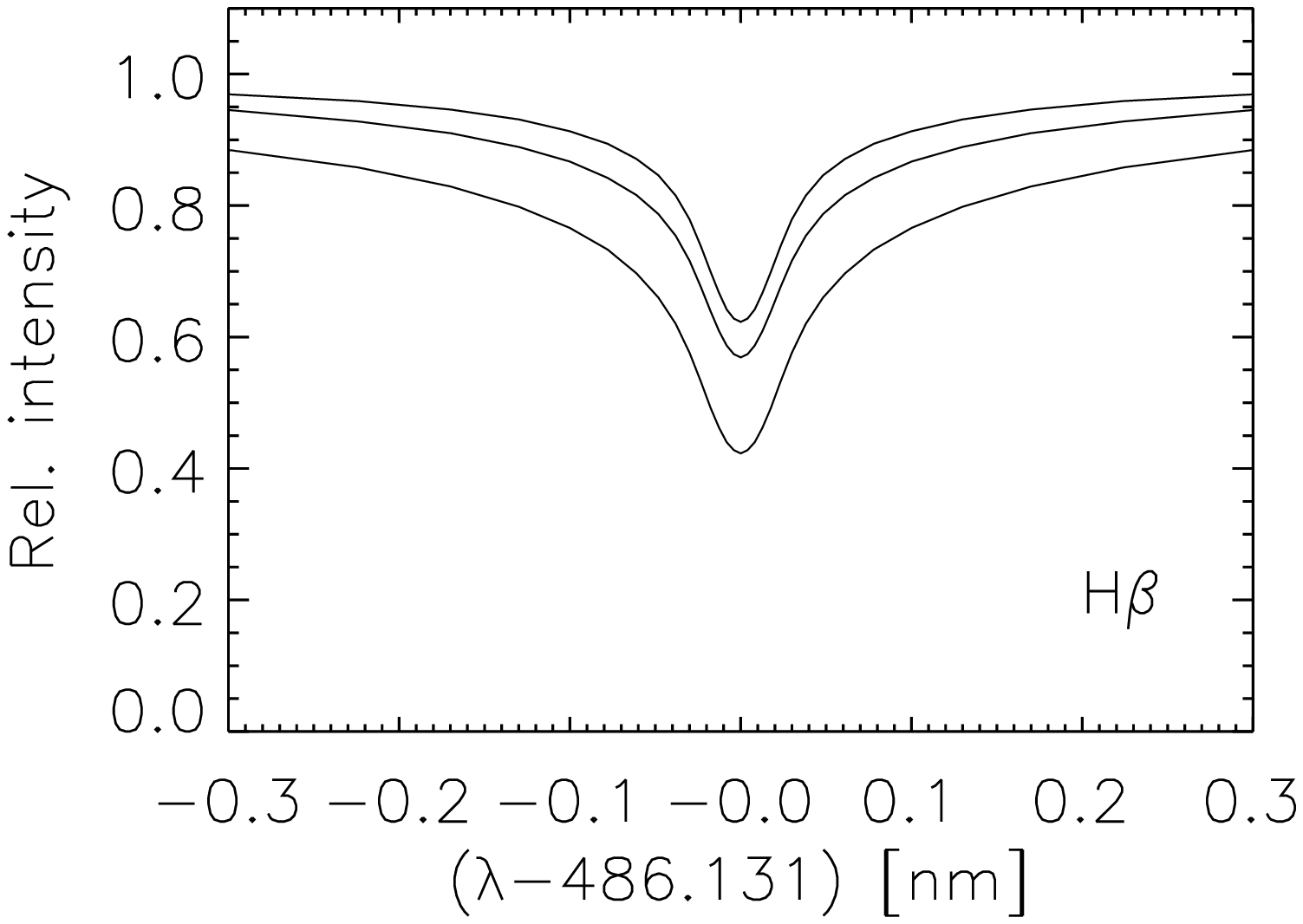}
\caption{NLTE line profiles from different positions on the stellar disk
in a red giant model with T=3750 K, $\log g$=0.5 . Left panel: profiles of
the hydrogen H$\alpha$ line from disk center at $\mu$=0.996 (lowest
curve), through $\mu$=0.5 (central curve at line core), to very near the
limb at $\mu$=0.004 (upper curve at line core). Here $\mu$ is the cosine
of the angle between the normal to the stellar surface and the line of
sight (referred to as ``the incidence angle'' hereafter). Right panel:
same for the hydrogen H$\beta$ line; the profile sequence is $\mu$=0.996
(lowest curve), $\mu$=0.004 (central curve) and $\mu$=0.5 (upper curve).
Note the difference in profile shape, $as~well~as$ in its change with
$\mu$.}
\label{fig:mimf2}
\end{figure}

Here we have limited our study to static (i.e., non-Mira) red giant
models. Our main interest is to compute realistic models of the
center-to-limb variation of the intensity on the stellar disk $-$ both for
individual spectral features in high resolution, as well as for the
optical pseudo-continuum. The state-of-the-art work for such models has
been done recently by Hofmann \& Scholz~(1998) and Jacob et~al.~(1999).
They employ model atmospheres from the grid of Bessell et~al.~(1989,
1991). Here we use similar models which reproduce their results. Striving
for this level of sophistication is crucial in our study of center-to-limb
and depth dependence of spectral features in red giants. For example,
Valls-Gabaud~(1998) used the linear limb darkening law and fitted simple
analytical expressions for spectral lines from Kurucz~(1992) model grids
to predict chromatic and spectroscopic signatures of microlensing events
of stars in general. We could not use the same simple approach, because we
find that for red giants these linear approximations fail by a factor
which is much larger than the uncertainties of the models, in line with
similar findings by Jacob et~al.~(1999).

The structure of the models in this paper is derived by assuming local
thermodynamic equilibrium (LTE) for the background opacities in
flux-constant plane-parallel atmospheres. We use the most recent opacity
sampling routines of Kurucz~(1999) and TiO line data of Schwenke~(1998).
These models serve as input for computations of hydrogen and calcium,
which are treated out of LTE (NLTE) with multi-level model atoms and in
spherical geometry (for more details see Loeb \& Sasselov~1995). The
hydrogen atom model has energy levels n=1$-$5, plus continuum; all
bound-bound and bound-free transitions are calculated in detail. The
calcium atom model contains 8 energy levels for the neutral species
(Ca{\sc I}) and 5 energy levels for the first ionization state (Ca{\sc
II}), plus continuum (Ca{\sc III}); all transitions are calculated in
detail. In the case of red giants, atomic hydrogen and calcium are not
important for the overall atmospheric structure. However, they still
provide strong spectral features with significant center-to-limb variation
(e.g. in the hydrogen Balmer lines, Ca{\sc II} H\&K, etc.). Most
moderately strong and weak lines diminish towards the limb (see
Figure~\ref{fig:mimf2}). As discussed further below, this variation could
be detected in a microlensing event by measuring the total equivalent
widths of spectral lines on medium-resolution spectra with a high
signal-to-noise ratio, or it could be seen directly on high-resolution
spectra (see \S 3).

\begin{figure}[tb]
\plottwo{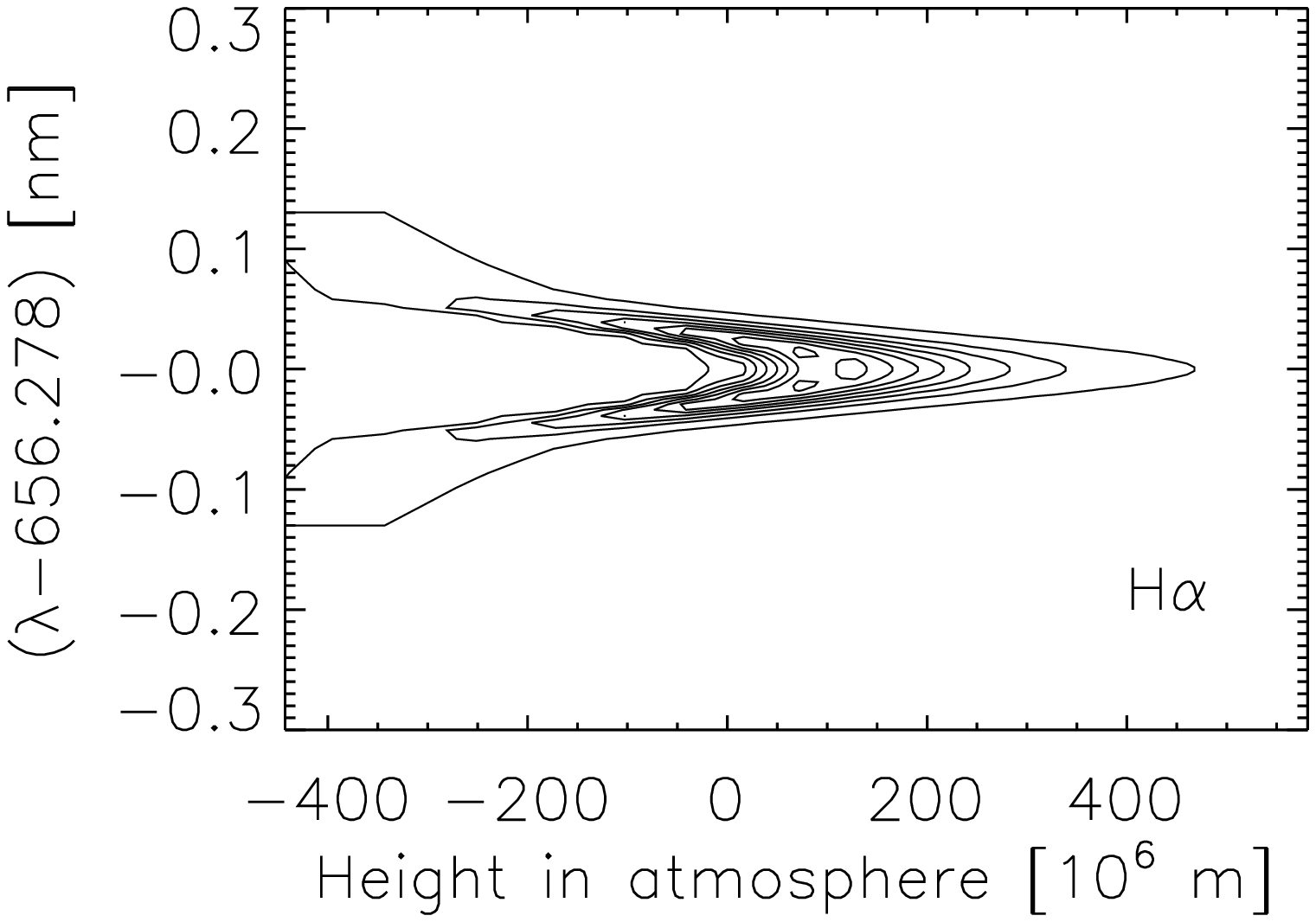}{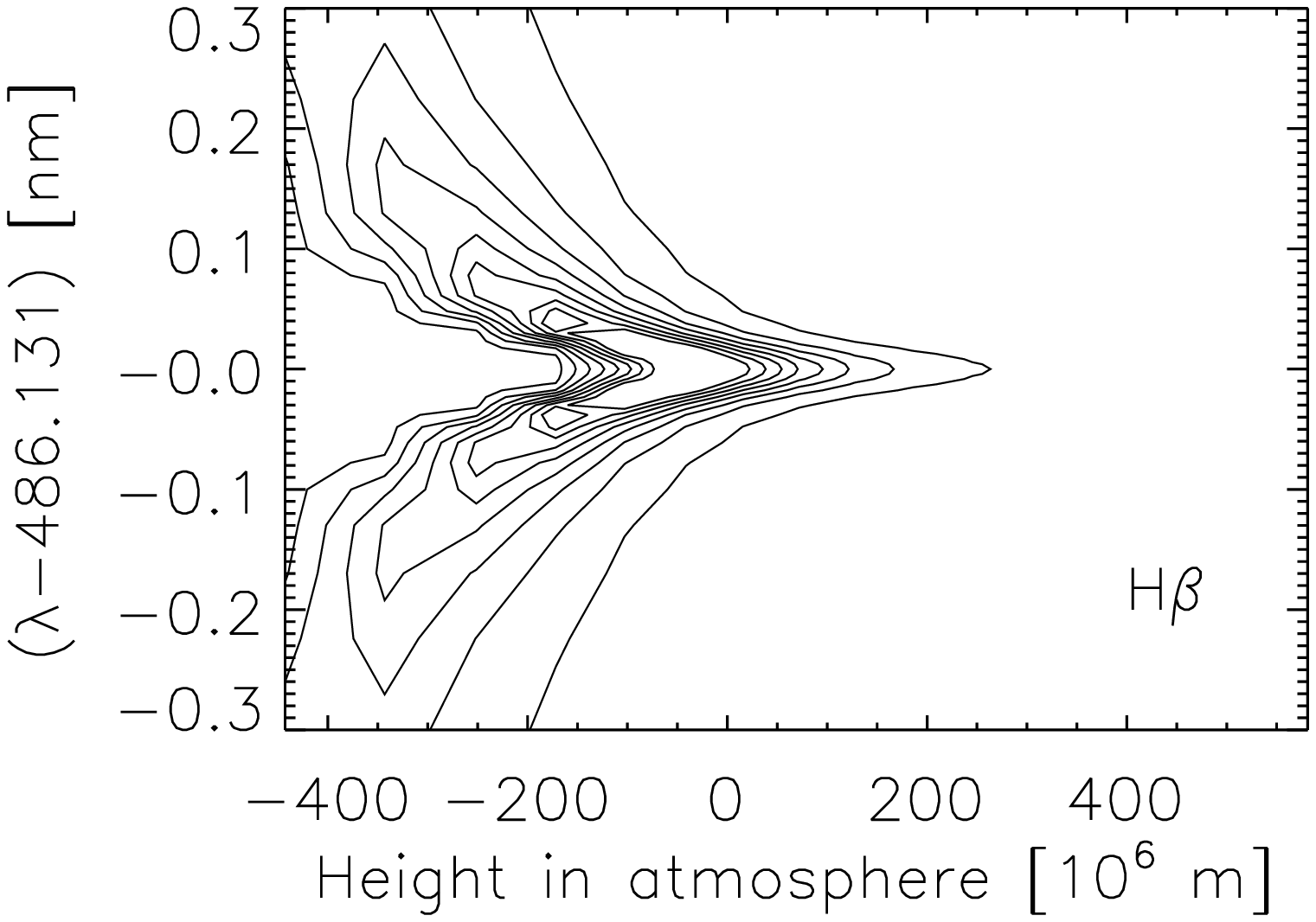}
\caption{Line-forming regions of the hydrogen H$\alpha$ and H$\beta$ lines
shown in Figure~\ref{fig:mimf2}, represented by their contribution
functions (CFs). The CF contours range from 0.1 (outermost) to 1.5
(innermost) in units of normalized intensity. Left panel: the hydrogen
H$\alpha$ line; right panel: the hydrogen H$\beta$ line. Zero height is
defined at continuum optical depth of unity for the rest wavelength of
each line respectively.}
\label{fig:figcf}
\end{figure}

The variation of the intensity from the center of the stellar disk to its
limb is studied theoretically by computing the contribution function of
each transition. This function defines the line-forming region in both
space (depth in the atmosphere) and frequency. The line-forming region is
shaped by local as well as non-local influences. Following Magain~(1986),
we define the NLTE contribution function (CF) for the relative line
depression as
$$
{CF(\log~\tau_{\lambda_0})~=~\frac{\ln10}{\mu}\,\tau_{\lambda_0}\, 
\frac{\kappa_l}{\kappa_0}\,\Bigl(1-\frac{S_l}{I_c}\,\Bigr)\, 
e^{-\tau_R /\mu}}, 
$$
where $\mu$ is the cosine of the incidence angle; $\tau_{\lambda_0}$ is
the optical depth at a reference wavelength ${\lambda}_0$\,; ${\tau}_R$ is
the optical depth corresponding to the opacity $\kappa_R = \kappa_l +
\kappa_c\,S_c/I_c$\,; $\kappa_0$ is the absorption coefficient at
${\lambda}_0$\,; $\kappa_l$ and $\kappa_c$ are the line and continuum
absorption coefficients, respectively; $S_l$ and $S_c$ are the line and
continuum source functions, respectively; and $I_c$ is the emergent
continuum intensity (if the line were absent). The traditional dependence
on frequency has been omitted from our notation for clarity. The CF for a
given synthesized spectral line indicates where the line is formed, and
thus it provides a theoretical tool for recovering the depth dependence of
thermodynamic variables in stellar atmospheres.

The CFs of the two lines presented in Figure~\ref{fig:mimf2} are shown in
Figure~\ref{fig:figcf}. Note how two spectral lines from the same element
and spectral series can exhibit differences in their CFs, in the CF's
location against the local continuum, and hence in their center-to-limb
variation. For example, the change in the strength of the H$\beta$ line
undergoes a reversal close to the limb of the star. The line gradually
weakens away from the disk center until it reaches a minimum, then it
becomes slightly more prominent again near the limb. In the case of the
H$\alpha$ line this effect is much weaker and occurs closer to the limb.
This difference between the two lines is due to the stronger dependence of
H$\beta$ on the TiO opacity, which is dominant at its wavelength.

\section {Effects of Microlensing on the Source Spectrum}

\subsection {Microlensing of an Extended Source}

Next we study the effect of microlensing by a point-mass on the spectrum
of the source star. For simplicity we describe the lensing configuration
in terms of angular distances in the plane of the sky, using the angular
source radius as a length unit.

In the absence of a lens, the observed flux from an unresolved source star
at wavelength~$\lambda$ is obtained by integrating the surface brightness
distribution $B(\lambda,\vec{r}\,)$ over the projected surface of the star
$\Sigma$,
\beq 
F_0(\lambda)=\int\limits_{\Sigma} B(\lambda,\vec{r}\,)\,d\Sigma\quad .
\label{eq:origflux} 
\eeq 
Any information about the surface structure is therefore concealed by the
integral. This information can be uncovered by microlensing, as its
signature contains differential information on the brightness
distribution. The flux from a simple point-source separated from a
point-mass lens by a projected distance $\sigma$ in the plane of the sky
gets amplified by a factor
\beq
A_{0}(\sigma)=\frac{\sigma^{2}+2\epsilon^{2}} 
{\sigma\sqrt{\sigma^{2}+4\epsilon^{2}}}\quad ,  
\label{eq:ptampl} 
\eeq 
where $\epsilon$ is the angular Einstein radius of the lens 
(Paczy\'nski~1996). The lensed flux from an extended source can then be
computed as
\beq
F(\lambda,\sigma_0)=\int\limits_{\Sigma} B(\lambda,\vec{r}\,) A_0(\sigma)\, 
d\Sigma\quad , 
\label{eq:lensflux} 
\eeq 
here $\sigma$ is the distance from the lens to point $\vec{r}$ on the
source, $\sigma_0$ is the distance to the source center. In our units,
$\sigma_0=1$ corresponds to having the lens projected at the source limb.  
The conversion between the $\sigma$ distances and the source-centered
position vector \mbox{$\vec{r}=(\rho\cos{\psi},\rho\sin{\psi})$} is
\beq
\sigma=\sqrt{\sigma_0^2-2\sigma_0\rho\cos{\psi}+\rho^2} \quad . 
\eeq 
The coordinates are oriented so that the lens lies in the direction
$\psi=0$. In this notation $\rho=\sqrt{1-\mu^2}$, where $\mu$ is the
cosine of the incidence angle used in \S 2. Generally the flux in
equation~(\ref{eq:lensflux}) depends also on the angle between the lens
and a fixed direction on the source (see Paper I). However, in this
work we consider for simplicity a circularly symmetric source with 
$B(\lambda,\vec{r}\,)=B(\lambda,\rho)$. Angular surface irregularities
such as spots are discussed in detail in a separate paper (Heyrovsk\'y
\& Sasselov~2000a). Equation~(\ref{eq:lensflux}) indicates that as the
lens moves, it gradually scans the source, giving highest weight to
the region of the star closest to the projected lens position. The
total resulting amplification of the source star is
\beq
A(\lambda,\sigma_0)=\frac{F(\lambda,\sigma_0)}{F_0(\lambda)} \quad .
\label{eq:ampl} 
\eeq

The center-to-limb surface brightness profile ($B$ as a function of
$\rho$) at different wavelengths has not only a different amplitude, but
also a different shape. As a consequence, the wavelength- and
position-dependence of the brightness distribution cannot be separated
$B(\lambda,\rho)\neq f_1(\lambda)f_2(\rho)$ and equation~(\ref{eq:ampl})
truly contains a chromatic dependence of the amplification. This fact does
not contradict the achromaticity of light deflection predicted by the
general theory of relativity, it merely reflects the different appearance
of the source star at different wavelengths.

For most microlensing events the point-source limit in
equation~(\ref{eq:ptampl}) gives a satisfactory description of the light
curve. Extended source effects become important only in events with a
sufficiently low impact parameter (defined as the projected distance of
closest approach of the lens and source on the sky). Expanding the
amplification~(\ref{eq:ampl}) in powers of the inverse of the lens-source
separation $\sigma_0^{-1}$ irrespective of the Einstein radius of the lens
(keeping the ratio $\epsilon/\sigma_0$ fixed) we obtain
\beq 
A(\lambda,\sigma_0)=\frac{\sigma_0^2+2\epsilon^2} 
{\sigma_0\sqrt{\sigma_0^2+4\epsilon^2}}  
\left[1+\frac{8\epsilon^4(\sigma_0^2+\epsilon^2)}
{\sigma_0^2(\sigma_0^2+2\epsilon^2)(\sigma_0^2+4\epsilon^2)^2}\,
\frac{\int\limits_0\limits^1 B(\lambda,\rho) \rho^3\, d\rho}
{\int\limits_0\limits^1 B(\lambda,\rho) \rho\, d\rho} + o\,(\sigma_0^{-2})
\right] \quad . 
\label{eq:ptlim} 
\eeq 
Strictly speaking, this $\sigma_0\gg1$ expansion is convergent for
$\sigma_0>1+\sqrt{2}$. The first extended source correction term in the
brackets is a product of two factors. The first one depends purely on the
lens parameters ($\epsilon, \sigma_0$), while the second one depends only on
the source properties ($B$). The lens factor is a monotonously decreasing
function of the lens distance, dropping from a value of $\sigma_0^{-2}/4$ when
the source lies well within the Einstein ring ($\sigma_0\ll\epsilon$) to a
value of $8\epsilon^4\sigma_0^{-6}$ when the source is well beyond the Einstein
ring ($\sigma_0\gg\epsilon$). The magnitude of the chromatic source factor is
less than one, typically on the order of $0.4$. Using this value we find
that any lens with $\epsilon\gtrsim4$ will produce a 1\% effect at
$\sigma_0=3$. Therefore, in order to achieve a larger extended source effect,
even a lens with high $\epsilon$ has to approach the source closer than 3
source radii, beyond the region where the above expansion is
valid\footnote{\, Lenses with $\epsilon\lesssim0.4$ are too weak to
trigger a microlensing alert in the current surveys. Lenses with 
$0.4<\epsilon<4$ require an approach closer than 3 source radii for a 1\%
effect.}. An observed example of such an event is M95-30 (Alcock 
et~al.~1997d), with a fitted value of the impact parameter
$p=0.715\pm0.003$ and $\epsilon=13.23\pm0.02$. 

Note that we estimate the extended source effect in comparison to a
point-source light curve with the same event parameters. Studying
detectability requires comparison with a best-fit point-source light
curve, which further lowers the upper limit for the closest approach.
Gould \& Welch~(1996) estimate that using color observations the 
finite-source effect is detectable out to 2 source radii.

\begin{figure}[tb]
\includegraphics{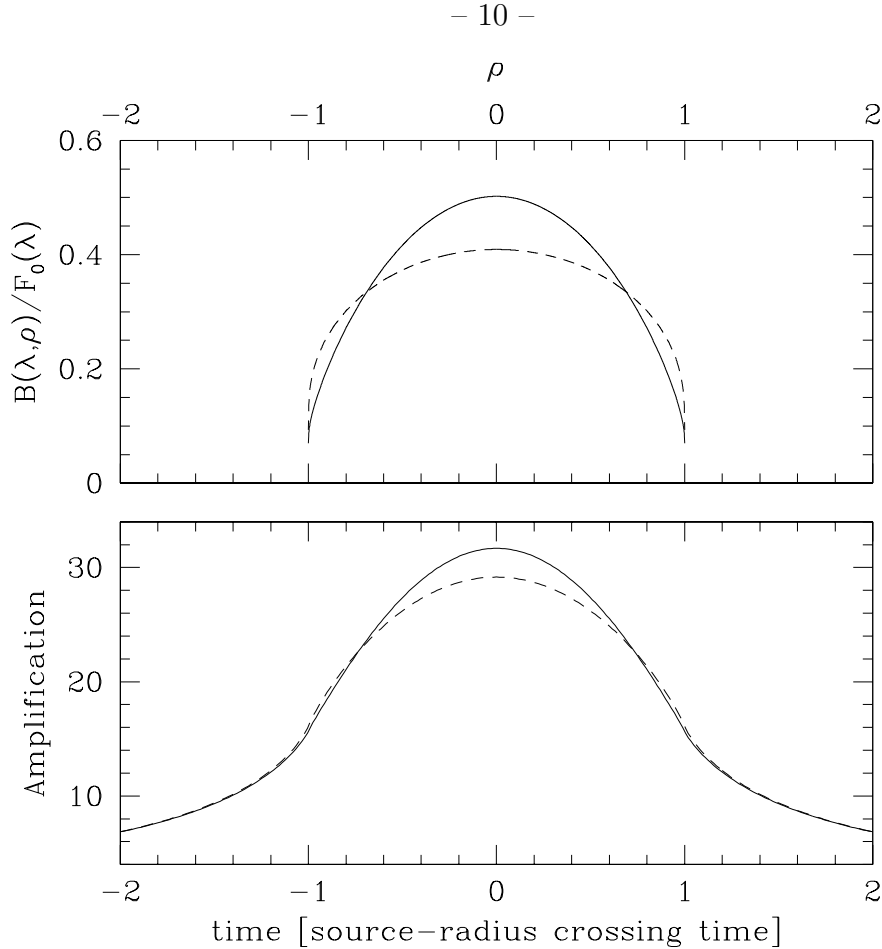}
\vspace*{4.5in}
\caption{Lower panel: light curves for two selected wavelengths (451~nm -
solid line, 759~nm - dashed line) of a T=3750 K, $\log g$=0.5 model
atmosphere (spectral resolution as in Figure~\ref{fig:amplpos}), lensed by
an $\epsilon=13.23$ lens with zero impact parameter. Upper panel:
corresponding normalized surface brightness profiles. For the purpose of
alignment with the lower panel the profiles are reflected to negative
values of $\rho$.}
\label{fig:2wav} 
\end{figure}

We study the case of close approaches or source transits directly from
equations~(\ref{eq:lensflux}) and (\ref{eq:ampl}). For evaluation of the
involved surface integral we follow the method described in Paper I. The
$\sigma^{-1}$ divergence of the amplification factor $A_0(\sigma)$ at the
point directly behind the lens can be avoided by integrating in
lens-centered polar coordinates. The product of $A_0(\sigma)$ with the
area element $d\Sigma=\sigma\,d\sigma\,d\phi$ then leads to a finite
integrand. For $B(\lambda,\rho)$ we use brightness profile data generated
for each wavelength directly from model atmosphere calculations (see~\S
2). We interpolate the computed center-to-limb data points by simple
quadratic segments $a+b\rho^2$. The computational advantage of this
approach is that the integral can then be analytically reduced to one
dimension (for details see Paper I). Additionally, as the lensed flux is
linear in the interpolation coefficients, it is sufficient to compute the
light curve for a single wavelength and then by substituting coefficients
obtain the other light curves for the same event. With this technique it
is feasible to perform the computation for a large number of wavelengths
and thus study spectral changes associated with microlensing.

\subsection {Light Curve Shapes and Low Resolution Spectra}

The effect of the brightness profile shape on the light curve is
demonstrated in Figure~\ref{fig:2wav}. For both selected wavelengths the
profile is normalized to unit total flux, in order to match the form in
which it appears in the amplification formula~(\ref{eq:ampl}). The light
curves plotted in the lower panel have a zero impact parameter, thus the
time $t$ measured from closest approach in units of source radius crossing
time is equal to the lens distance ($|\,t\,|=\sigma_0$). For illustration
we use the M95-30 lens with $\epsilon=13.23$. From the figure it can be
seen that the light curve traces the brightness profile shape - the
flatter profile produces a flatter light curve. In addition, the light
curve with lower amplification at the source center typically crosses the
other at $\sigma_0\sim 0.7$ and has a higher amplification at the limb of
the source. We note that flatter profiles usually indicate formation
higher in the stellar atmosphere.

\begin{figure}[tb]
\includegraphics{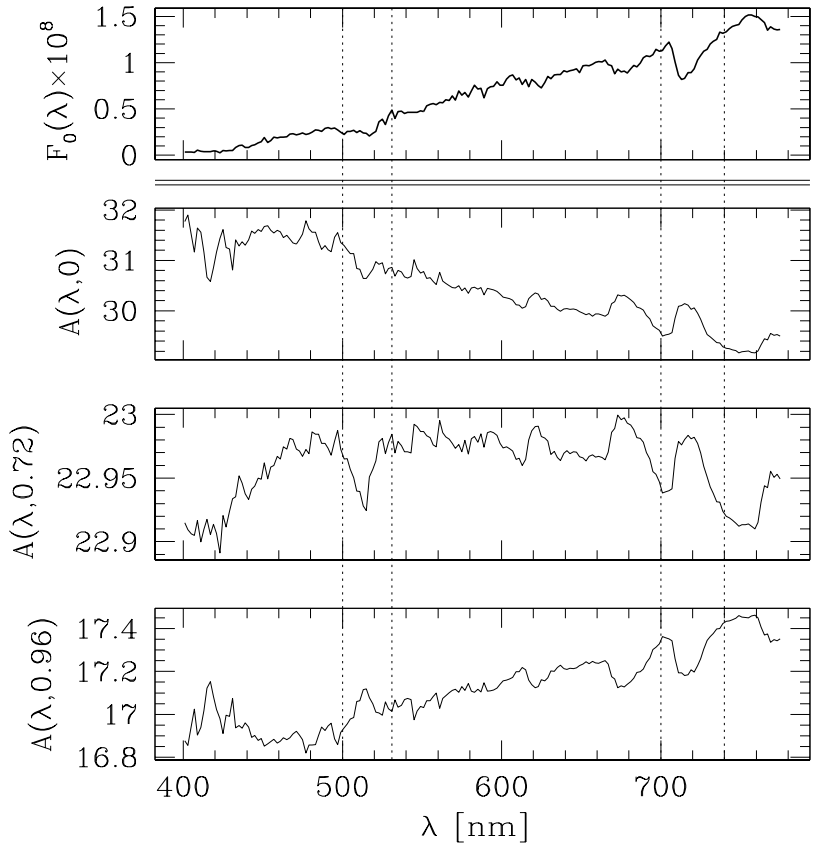}
\vspace*{4.5in}
\caption{Upper panel: low resolution unlensed spectrum of the T=3750 K,
$\log g$=0.5 model atmosphere (flux in units of $W\ m^{-2}\ Hz^{-1}$).
Lower three panels: amplification by an $\epsilon$ = 13.23 lens as a
function of wavelength for lens positions $\sigma_0$ = 0, 0.72 and 0.96.
Vertical dotted lines mark two regions with TiO bands affected by 
microlensing in opposite ways. These regions are enlarged with high
spectral resolution in Figure~\ref{fig:500700}.}
\label{fig:amplpos} 
\end{figure}

Figure~\ref{fig:amplpos} illustrates the wavelength dependence of the
amplification on a low resolution (R=500) spectrum from 401 to 775~nm.  
These results are obtained using the same model atmosphere as in
Figure~\ref{fig:2wav}, with T=3750 K and $\log{g}=0.5$, lensed by an
$\epsilon=13.23$ lens. The model spectrum in the absence of the lens is
shown in the upper panel. The lower three amplification plots correspond
to lens positions $\sigma_0=0$ (source center), 0.72 (closest approach of
M95-30) and 0.96 (close to the limb). Comparison of the plots at the
center and the limb further illustrates the relative change in
amplification at different wavelengths seen in Figure~\ref{fig:2wav} $-$
the amplification curve at the limb is virtually a vertical flip of the
curve at the center. The overall slope of the curves indicates that for
$\sigma_0=0$ the spectrum will appear bluer and for $\sigma_0=0.96$ redder
than in the absence of a lens (see also Gould \& Welch~1996; 
Valls-Gabaud~1998). 

The finer structure of the curves provides further interesting results.  
Sections of the curves with significant spectral variations of
amplification indicate spectral regions particularly sensitive to
microlensing. Observations of such regions during a transit event can then
be used to test the depth dependence of the atmosphere model. While the
resolution of this particular spectrum is too low to resolve individual
spectral lines, broader molecular bands are clearly visible. An example of
a sensitive feature is the TiO band system at 710~nm ($A\!-\!X$,
$\gamma$-bands), as hinted by the bump in the amplification curve. Recent
models by Jacob et~al.~(1999) also point to the extreme limb-darkening
effects in this band system. It should be noted that not all bands are
affected in the same manner $-$ the $C\!-\!X$ TiO system at 516~nm
($\alpha$-bands) shows an opposite variation in amplification. The nature
of the change in these bands is discussed in more detail further below.

\begin{figure}[tb]
\includegraphics{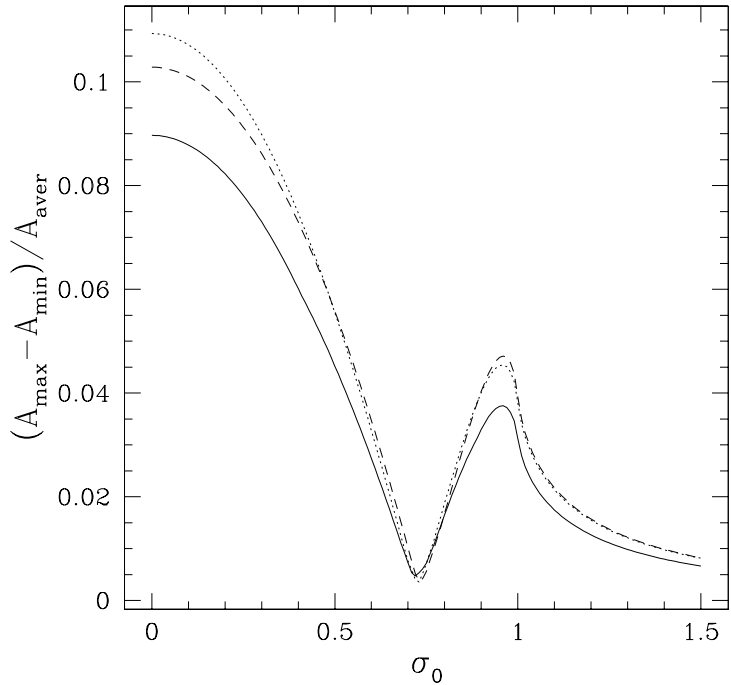}
\vspace*{3.7in}
\caption{Variation of amplification within the low resolution spectrum as
a function of lens position for three model atmospheres: T=3750 K, $\log
g$=0.5 (solid line), T=3500 K, $\log g$=1 (dashed line) and T=4000 K,
$\log g$=0.5 (dotted line). The Einstein radius of the lens is the same 
as in previous figures ($\epsilon$ = 13.23).} 
\label{fig:amplmod} 
\end{figure}

The three amplification curves are each plotted for convenience with a
different vertical scale. For example, the relative amplification
variation at $\sigma_0=0.72$ is much smaller than at either the center or
the limb. A simple way to quantify the degree of this variation is to take
the maximum, minimum and average amplification over the full range of
wavelengths and plot the ratio $(A_{max}-A_{min})/A_{aver}$ as a function
of lens position $\sigma_0$. This ratio in fact serves as a measure of
chromaticity - it drops to zero in the achromatic point-source limit.
Results for three different atmosphere models (wavelength range
401--775~nm) are presented in Figure~\ref{fig:amplmod}. While there is a
difference in overall amplitude, the general character of all three curves
is similar. As the lens approaches the source, the degree of chromaticity
increases, reaching a local peak of~$\sim$~3--5\% just after crossing the
limb. If the lens passes closer to the center, the chromaticity drops
to~$<$~0.5\% at $\sigma_0\sim 0.7$ (approximately the M95-30 impact
parameter, by coincidence). This indicates that the light curves for all
the studied wavelengths intersect approximately at the same point, as seen
in Figure~\ref{fig:2wav}. A yet closer approach causes the chromaticity to
increase to a peak value of~$\sim$~8--11\% when the lens is positioned at
the source center. To summarize, the strongest spectral effects can be
expected when the lens is aligned with the source center, with weaker
effects near the limb. In between these two positions there is a drop in
chromaticity nearly to zero (achromaticity). Although the limb effects are
weaker, in principle they should be observable in any transit event.

\begin{figure}[tb]
\includegraphics{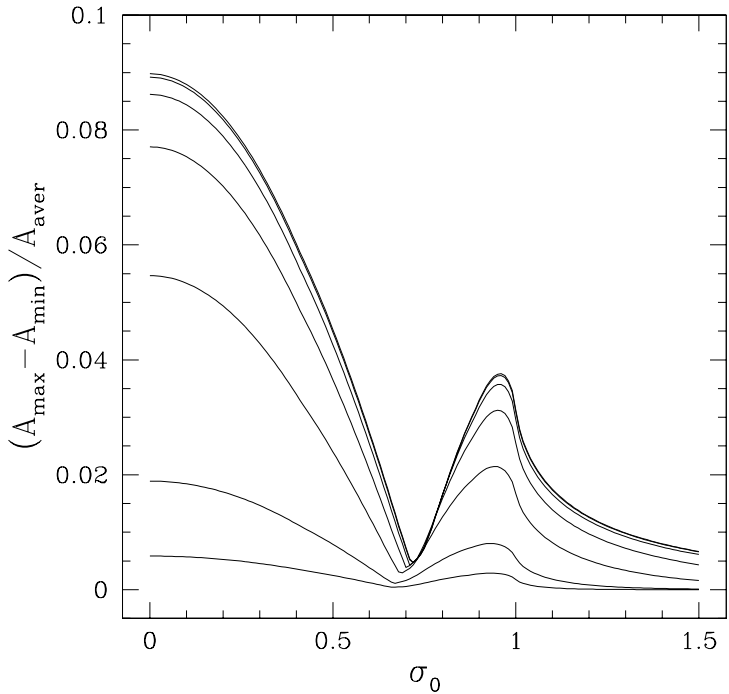}
\vspace*{3.7in}
\caption{Variation of amplification within the low resolution spectrum 
(401--775~nm) as a function of lens position for the T=3750 K, $\log
g$=0.5 model atmosphere. Curves correspond to lenses with different Einstein 
radii in units of the source radius (from lowest curve): $\epsilon$ = 
0.1, 0.2, 0.5, 1, 2, 5, 100.} 
\label{fig:ampleps}
\end{figure}

The results in Figure~\ref{fig:amplmod} were computed using the M95-30
lens with $\epsilon=13.23$. The dependence of the results on the Einstein
radius is explored in Figure~\ref{fig:ampleps} for the model atmosphere
with T=3750 K and $\log{g}=0.5$. While the chromaticity predictably
increases with $\epsilon$, its $\sigma_0$-dependence rapidly reaches a limit
curve. In other words, the degree of lensing chromaticity cannot increase
beyond this curve for a given atmosphere model and fixed spectral
resolution. This behavior can be understood by studying the form of
equation~(\ref{eq:ampl}) during a close approach (as compared to the
Einstein radius). In the limit\footnote{\,Convergence requires also
$\epsilon>(\sigma_0+1)/2\,$.} $\epsilon\gg \sigma_0$ the amplification
\beq 
A(\lambda,\sigma_0)=\epsilon\ 
\frac{\int\limits_{\Sigma}\sigma^{-1} B(\lambda,\rho)\,d\Sigma}
{\int\limits_{\Sigma}B(\lambda,\rho)\,d\Sigma}\,+
O\left(\frac{\sigma_0}{\epsilon}\right) 
\label{eq:amplim}
\eeq 
is directly proportional to the Einstein radius $\epsilon$. The degree of
chromaticity depends on amplification ratios and is therefore independent
of $\epsilon$ in this limit, as seen in Figure~\ref{fig:ampleps}. We
conclude that as far as spectral effects are concerned, the ``only''
advantage of microlensing by stronger lenses ($\epsilon>5$) is an overall
increase in flux. 

\begin{figure}[tb] 
\includegraphics{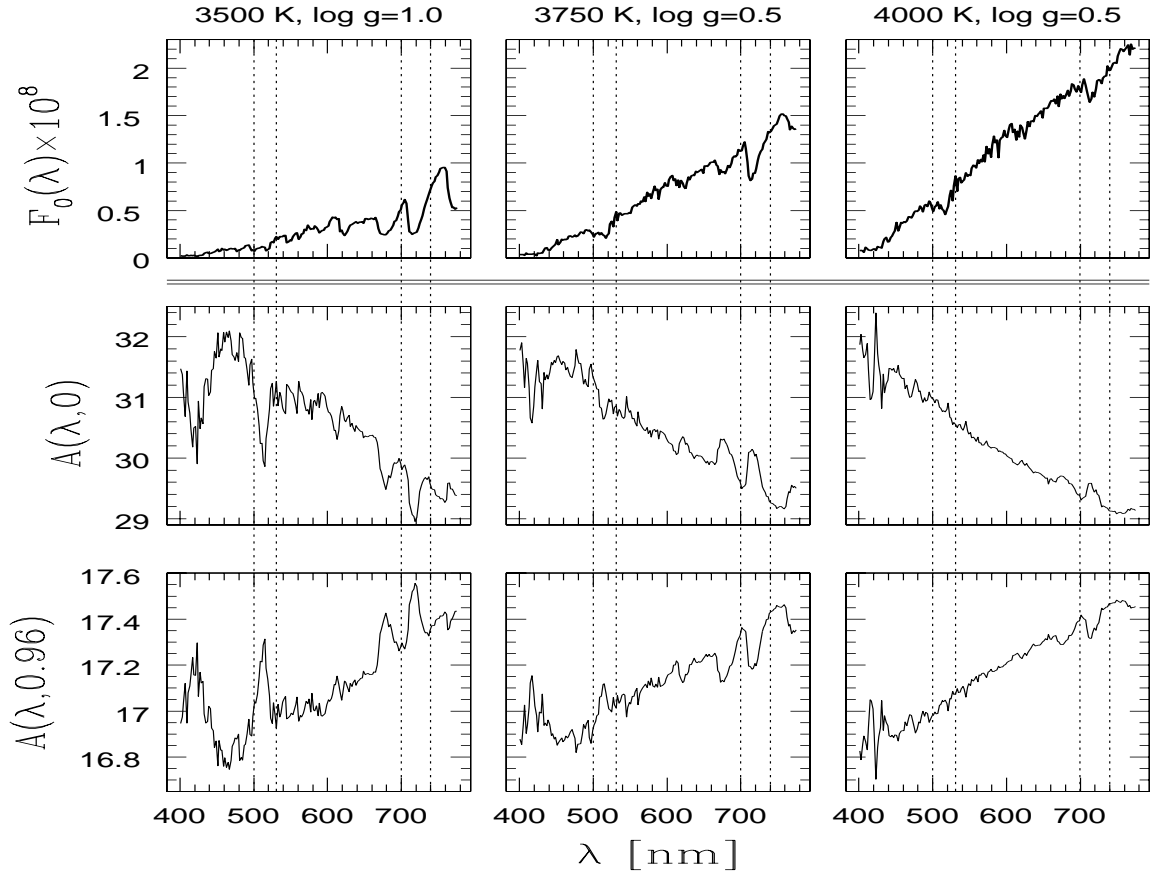}
\vspace*{4.45in}
\caption{Microlensing by an $\epsilon$ = 13.23 lens of three different
model atmospheres: T=3500~K, $\log g$=1.0 (left column); T=3750~K, $\log
g$=0.5 (central column) and T=4000~K, $\log g$=0.5 (right column). Upper
row: low resolution unlensed spectra (flux in units of $W\ m^{-2}\ 
Hz^{-1}$); central row:  amplification for lens position $\sigma_0$ = 0; lower
row: amplification for lens position $\sigma_0$ = 0.96. Vertical dotted lines
mark the TiO $\alpha$-bands (516~nm) and $\gamma$-bands (710~nm).}
\label{fig:ampldiff}
\end{figure}

Figure~\ref{fig:amplpos} illustrates the capability of microlensing to
resolve the depth structure of a given stellar atmosphere. Can the
microlensing signature also be used to distinguish between different
atmospheres? Figure~\ref{fig:ampldiff} contains the results of
Figure~\ref{fig:amplpos} together with those computed for the two other
atmosphere models used in Figure~\ref{fig:amplmod}; a cooler one
(T=3500~K, $\log g$=1) and a hotter one (T=4000~K, $\log g$=0.5). As
expected, the results indeed vary substantially between the different
models. In the 3500~K atmosphere, for example, the TiO $\alpha$-bands at
516~nm show a stronger amplification variation, and more importantly, the
TiO $\gamma$-bands at 710~nm show the opposite variation than in the
3750~K case. The amplification variation in the 4000~K model is much
weaker overall, because the TiO bands are less prominent at this higher
temperature. The only prominent amplification feature at 420~nm is in a
spectral region with very low flux, hence it would be difficult to detect
any spectral changes there.

\begin{figure}[tb] 
\plottwo{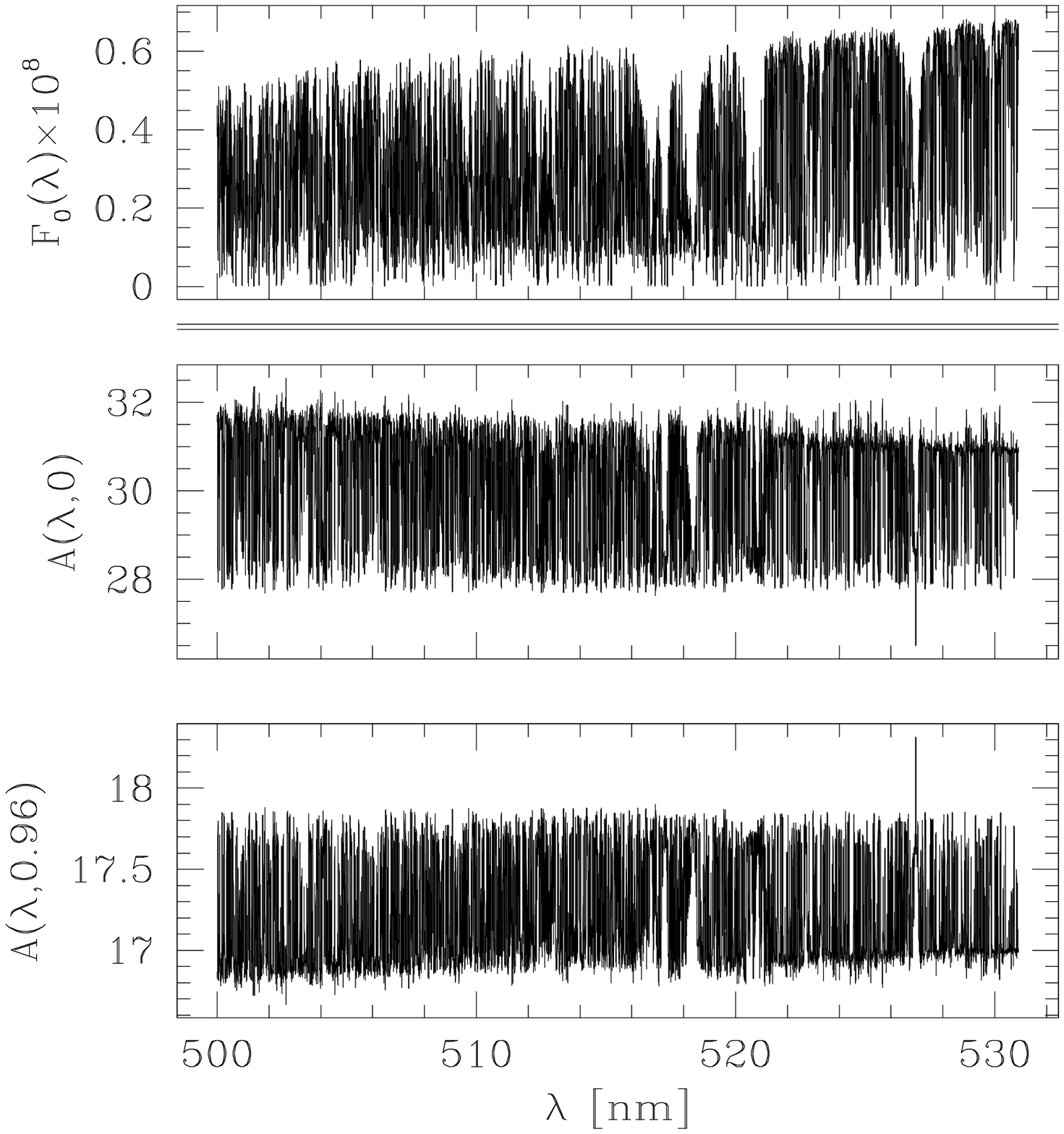}{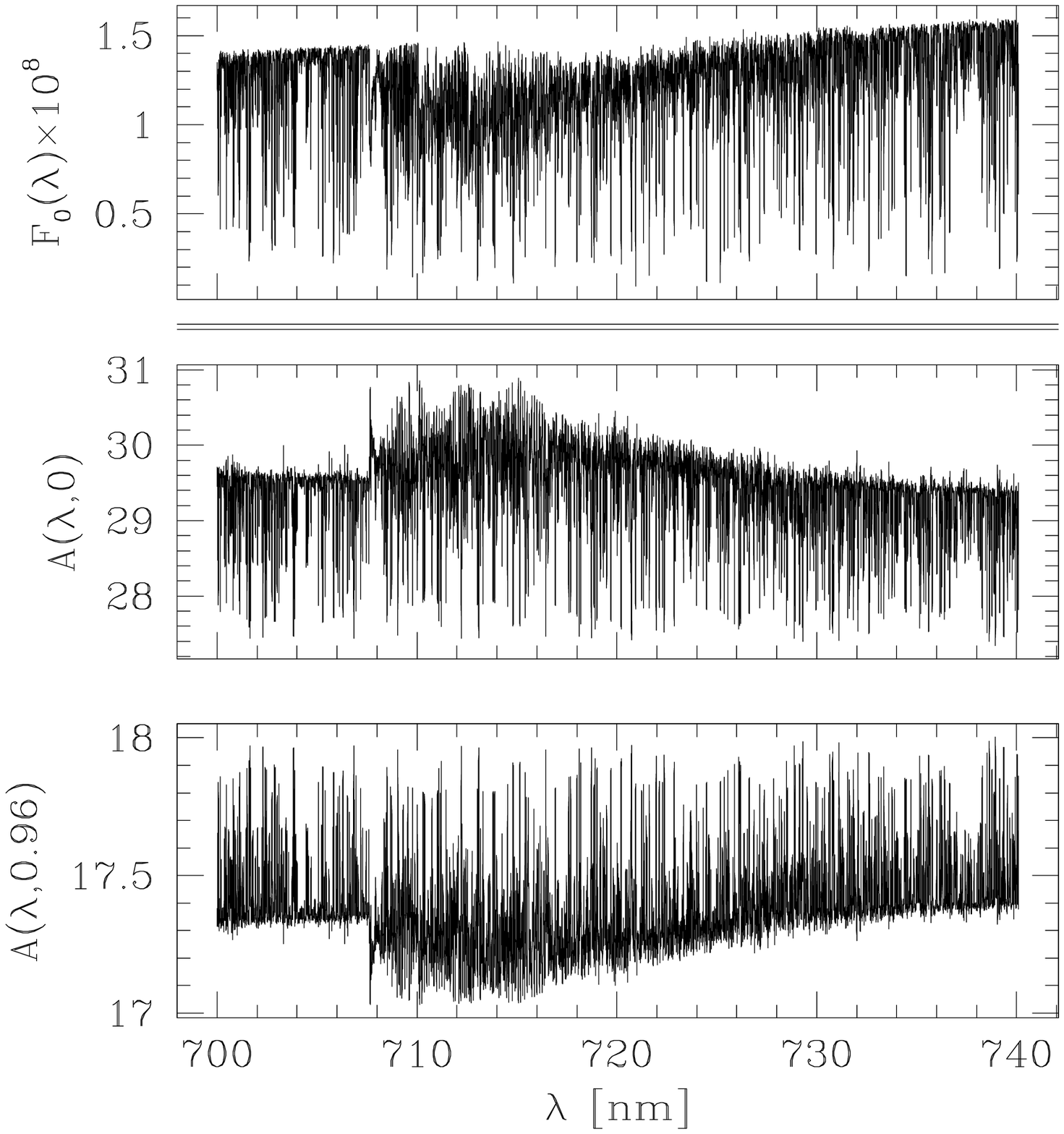}
\caption{Upper row: high resolution unlensed spectrum of the T=3750~K,
$\log g$=0.5 model atmosphere (flux in units of $W\ m^{-2}\ Hz^{-1}$).
Lower two rows:  amplification by an $\epsilon$ = 13.23 lens as a function
of wavelength for lens positions $\sigma_0$~=~0 and 0.96. Left panels: spectral
region near the 516~nm TiO $\alpha$-bands. Right panels: spectral region
near the 710~nm TiO $\gamma$-bands.}
\label{fig:500700}
\end{figure}

\subsection {High Resolution Spectra and Line Profiles}

The results presented in Figures~\ref{fig:amplpos}, \ref{fig:amplmod},
\ref{fig:ampleps} and \ref{fig:ampldiff} illustrate the overall
microlensing effects on low resolution spectra. However, studying the
behavior of finer features, such as individual molecular bands or
spectral lines, can provide stronger constraints on model atmospheres.
First we return to the TiO bands mentioned above. In this particular
case, even the low resolution results illustrated in
Figure~\ref{fig:amplpos} indicate the different behavior of the two band
systems. When the lens is at $\sigma_0=0$, the $\gamma$-band system at
710~nm has a higher amplification in its region of highest absorption.
As a result, the band system will appear weaker (shallower) than in the
absence of the lens. With the lens at the limb, the highest absorption
region has a lower amplification than the adjacent spectral region. The
band system will therefore be more prominent than if the lens were
absent. In contrast, the $\alpha$-band system at 516~nm shows exactly
the opposite behavior. It will be more prominent with the lens at the
source center, and less prominent when the lens is at the limb. The
different behavior of the $\gamma$-band system is due to the extension
of its region of formation higher in the atmosphere.

\begin{figure}[tb]
\includegraphics{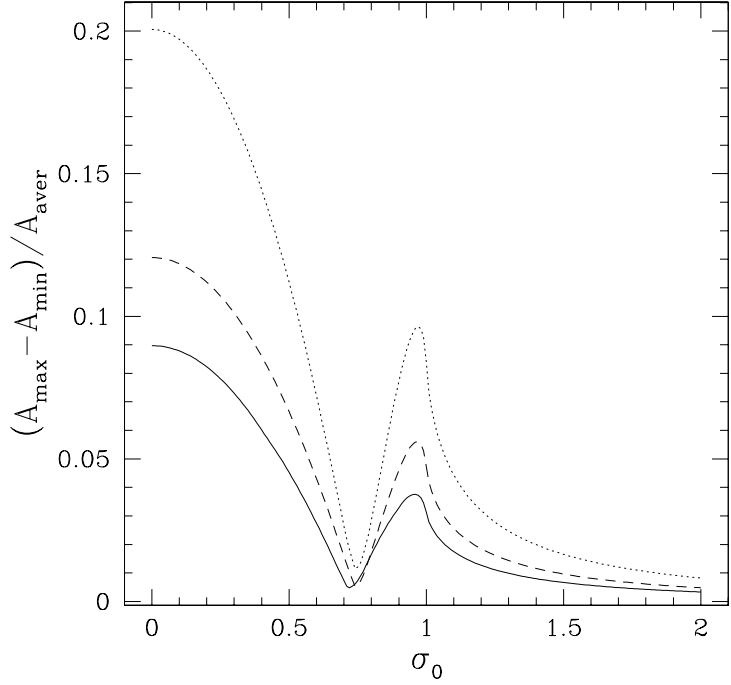}
\vspace*{3.7in}
\caption{Variation of amplification as a function of lens position for the
T=3750 K, $\log g$=0.5 model atmosphere lensed by an $\epsilon$ = 13.23
lens. Low resolution spectrum 401-775~nm (solid line), high resolution
500-531~nm (dotted line) and high resolution 700-740~nm (dashed line).}
\label{fig:amplhilo}
\end{figure}

These effects are verified by the results of high spectral resolution
(R=500,000) calculations shown in Figure~\ref{fig:500700}. The first
feature to notice in these plots, apart from the confirmed band system
structure, is the high ``fuzziness'' of the amplification curves. This
property indicates that the amplification varies highly within individual
spectral lines. In fact the less ``fuzzy'' 710~nm region also has a lower
density of spectral lines. Studying the amplification variation of both
regions in Figure~\ref{fig:amplhilo}, it comes as no surprise that while
the shape of the dependence is similar, the degree of chromaticity is
higher than in the full extent of the low resolution spectrum. This higher
variation is due to the fine spectral structure, which is averaged out in
the low resolution results. The degree of chromaticity for the 516~nm and
710~nm regions in high resolution again peaks at the source center (20\%
and 12\%, respectively) and has a secondary peak at the limb (9.5\% and
5.5\%, respectively). For comparison, the center and limb values for the
low resolution spectrum are 9\% and 4\%, respectively. Note that the high
values for the 516~nm region are partly caused by the single atomic
spectral line at 526.95~nm \footnote{\,Unfortunately this line is highly
saturated at its core, where the amplification undergoes the high
variation - the effect will therefore be difficult to detect because of
the low photon flux at the line center.}.

\begin{figure}[tb]
\includegraphics{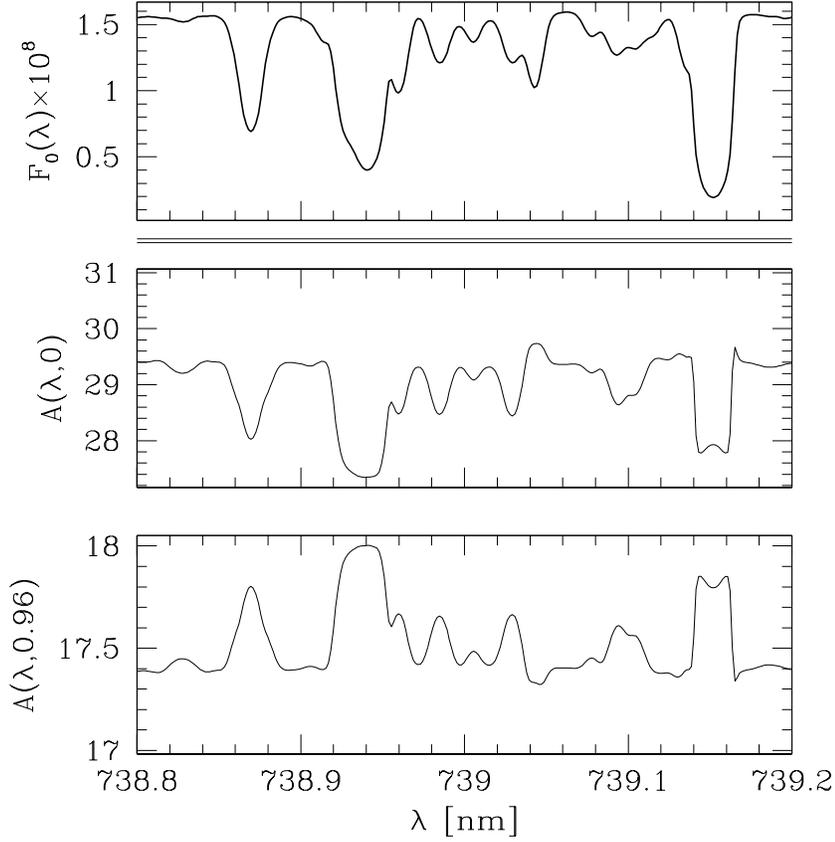}
\vspace*{4.5in}
\caption{Detail of Figure~\ref{fig:500700} resolving individual spectral
lines. The spectral region shown is between 738.8--739.2~nm.}
\label{fig:blowup}
\end{figure}

A closer look at the fine structure in both sets of high resolution
results (apart from the large-scale TiO band effect) suggests a
significant correlation between the unlensed spectrum and the
amplification curve for $\sigma_0=0$. A blow-up of a 0.4~nm interval in
Figure~\ref{fig:blowup} confirms the correlation - most individual lines
have their cores amplified less than their wings at $\sigma_0=0$, while at
$\sigma_0=0.96$ the cores are amplified more than the wings. Hence most
lines become more prominent with the lens at the source center, and less
prominent with the lens close to the limb than in the absence of the lens
($\sigma_0\rightarrow\infty$). There are fewer examples of lines with the
opposite behavior (e.g. at 739.044~nm in Figure~\ref{fig:blowup}), with
virtually no change (flat amplification curve) or with more complex
behavior.

The characteristic correlation can be explained for typical absorption
lines in the following way. Such lines form at a certain depth in the
atmosphere, and gradually become less prominent away from the center
towards the limb, as there is relatively less absorption at the limb. The
center-to-limb variation curves at the cores of these lines are therefore
flatter than in the adjacent continuum. The correlation is now directly
explained by the discussion of light curve shapes above - when the lens is
at the source center, the line cores are amplified less than the
continuum, while with the lens at the limb the cores are amplified more
than the continuum.

The less common anti-correlation (or lack of correlation) of the unlensed
spectrum and the amplification at source center can be expected for
example in some emission features, high excitation lines, or in spectral
lines affected by more complex physical effects, such as resonant
scattering (Loeb \& Sasselov~1995), overlapping lines, lines within
molecular bands, etc.

\begin{figure}[tb]
\plottwo{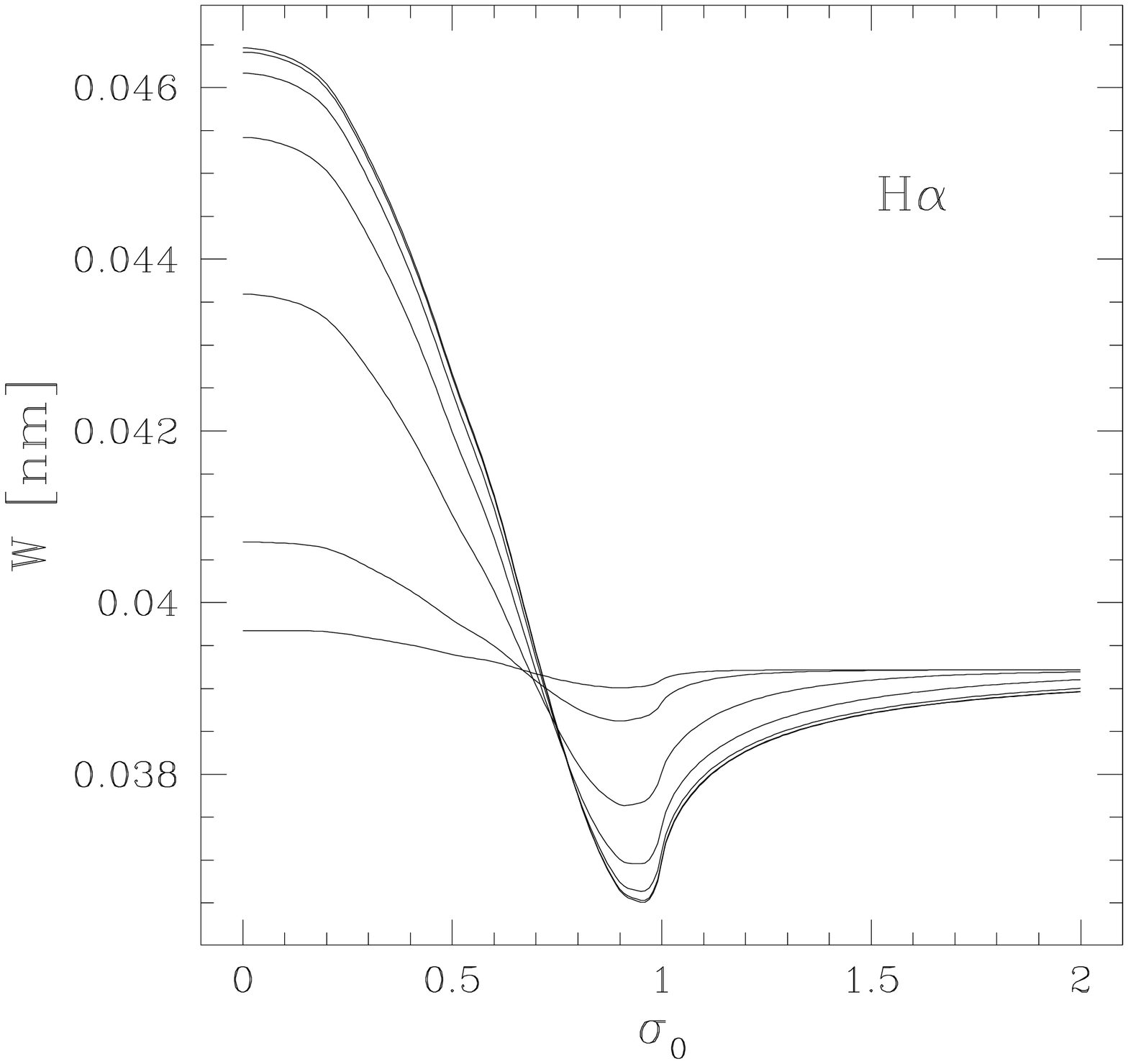}{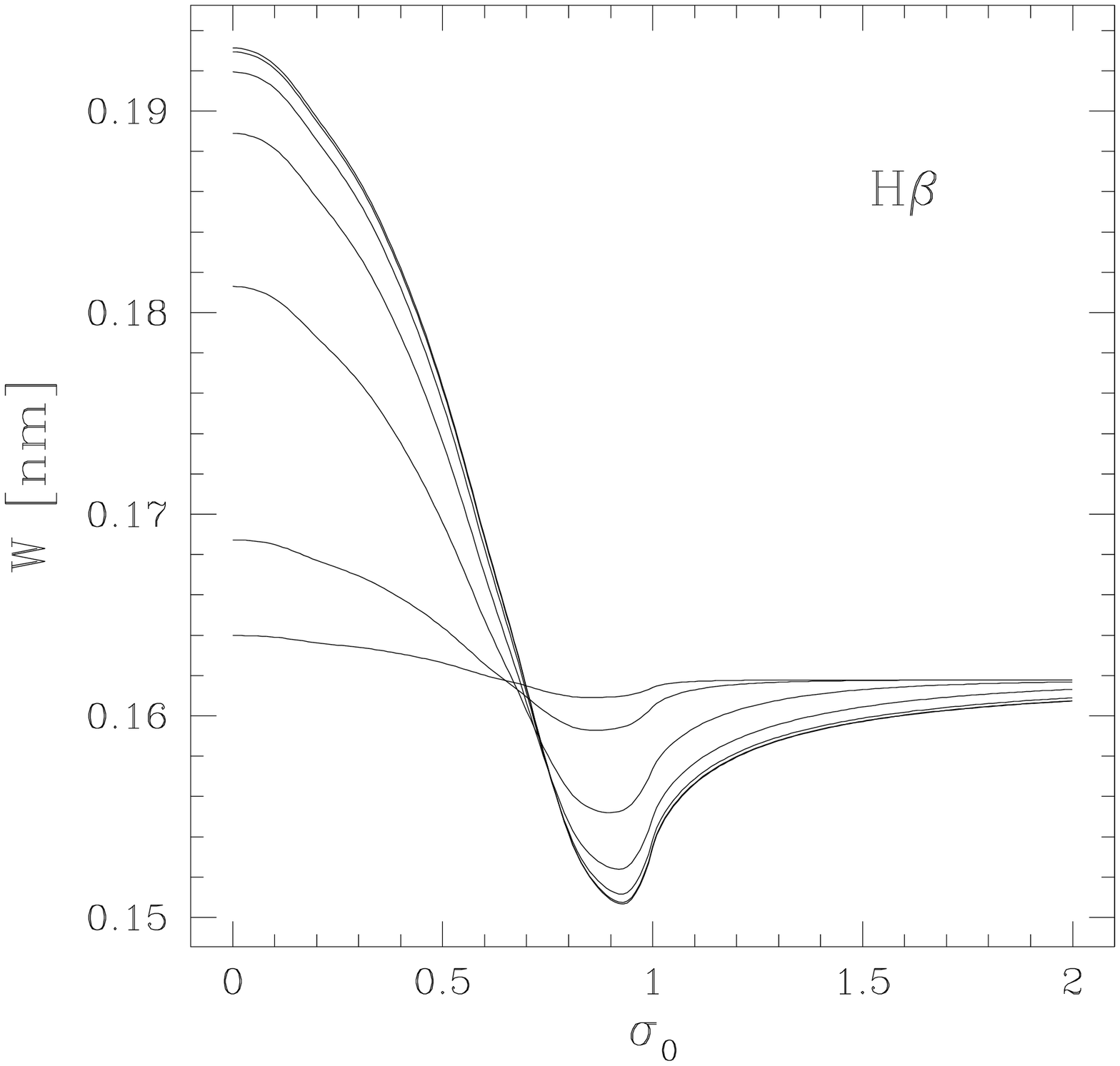}
\caption{Equivalent width of H$\alpha$ (left panel) and H$\beta$ (right 
panel) as a function of lens position, for the T=3750 K, $\log g$=0.5
model atmosphere. Individual curves correspond to different lenses (from
lowest at $\sigma_0$ = 0): $\epsilon$ = 0.1, 0.2, 0.5, 1, 2, 5, 100.}
\label{fig:halhbe}
\end{figure}

A straightforward method for studying the change of a particular line is
to compute its equivalent width, which is generally given by 
\beq
W(\sigma_0)=\Delta\left[ 1-\frac{\int F(\lambda,\sigma_0)\,d\lambda} {\int 
F_c(\lambda,\sigma_0)\,d\lambda} \right] \quad ,
\label{eq:width}
\eeq
where $F_c(\lambda,\sigma_0)$ is the continuum flux and the integrals are
performed over a wavelength interval of width $\Delta$ around the line. As
an example we present here results for the hydrogen Balmer lines H$\alpha$
and H$\beta$ in the same T=3750 K, $\log g$=0.5 model. The center-to-limb
variation of their unlensed profiles is shown in Figure~\ref{fig:mimf2}.
The dependence of their equivalent widths on lens position is plotted in
Figure~\ref{fig:halhbe} for a set of Einstein radii $\epsilon$. First we
note that the change with $\epsilon$ has the same character as in
Figure~\ref{fig:ampleps}. Any sufficiently strong lens ($\epsilon>5$) will
therefore affect the equivalent width in the same way. This result can
again be traced to the fact that the equivalent width in
equation~(\ref{eq:width}) depends on a ratio of amplifications. This ratio
is independent of $\epsilon$ in the $\epsilon\gg \sigma_0$ limit, as seen
from equation~(\ref{eq:amplim}). During a microlensing event both Balmer
lines behave in the typical manner described above. An approach of a
sufficiently strong lens first causes the width of both lines to drop (by
$\sim$7\%), as it gives higher weight to the limb where the lines are
weak. Similarly, a lens positioned at the center, where both lines are
strong, would increase their width ($\sim$19\%) above the asymptotic
value. Both the H$\alpha$ and H$\beta$ curves are similar, although those
of H$\beta$ are more centrally peaked and have a broader minimum near the
limb.

We compared our results to those of Valls-Gabaud~(1998) for H$\beta$ with
the same event parameters\footnote{\, We neglect here the earlier results
in Valls-Gabaud~(1996), which are inconsistent with Valls-Gabaud~(1998).}.  
The general character of the change with lens position is the same -
decrease of the equivalent width (EW) at the limb and a higher increase
closer to the source center. However, Valls-Gabaud's results show a two to
three times lower relative change of the EW, the actual factor depending
on the lens position. As a result, the predicted ratio of EW increase
(close to the center) to peak EW decrease (at the limb) is also lower than
our computations show. Furthermore, the achromaticity point appears to be
closer to the limb (around $\sigma_0=0.8$) than in our analysis. As
mentioned in \S 2, assumptions such as linear limb darkening degrade
Valls-Gabaud's models too much to give realistic predictions of equivalent
widths, particularly in the case of red giants. This comparison further
illustrates the sensitivity of microlensing to the atmosphere structure of
the resolved source.

A detailed study of high resolution spectral changes in different model
atmospheres and comparison with the M95-30 observational results will be
presented elsewhere (Heyrovsk\'y \& Sasselov~2000b).

\section{Conclusions}

Microlensing events in which a point-mass lens with Einstein radius
$\epsilon>4$ (in source radius units) approaches the source star within
three source radii already exhibit $\sim$1\% deviations from
point-source light curves. The deviations further increase for closer
approaches. These finite source effects depend on the surface brightness
distribution of the source star. Observations of such effects,
particularly during source transit events, can therefore be used to
resolve and study the structure of the stellar disk. As the brightness
distribution depends on wavelength, the usual assumption of microlensing
achromaticity breaks down in these cases. Spectral observations can then
be used to probe the depth structure of the atmosphere of the source
star. The most promising sources for such studies are red giants in the
Galactic bulge, among other reasons due to their large size, intrinsic
brightness and fairly high abundance. These circumstances are in fact
particularly fortuitous, as the structure of cool red giant atmospheres
is currently very poorly observationally constrained.

The degree of microlensing chromaticity can be measured by the relative
spectral variation of the microlensing amplification. Thus defined,
chromaticity as a function of lens position is found to have two peaks -
primary peak at the source center and secondary peak close to the source
limb, with a dip to achromaticity in between. In the usually observed type
of events, in which the Einstein radius $\epsilon>5$, chromaticity is
independent of $\epsilon$. Its peak values for a low resolution optical
spectrum are typically 10\% at the center and 4\% at the limb (both values
increase with higher spectral resolution). Any such transit event should
therefore exhibit at least the 4\% effect, which should be readily
observable under favorable conditions. In future potential source-crossing
events it is therefore important not to miss the time when the lens is
close to the limb, as was unfortunately the case in M95-30.

During a red giant transit event the overall spectrum appears redder when
the lens is at the source limb, and bluer if the lens comes close to the
source center. Individual spectral lines respond to microlensing in
different ways, because of their different center-to-limb variation, which
reflects their different depth of formation in the star's atmosphere. Most
absorption lines will typically turn weaker (lower equivalent width) when
the lens is close to the limb and more prominent when the lens is near the
center. Some emission lines, high excitation lines, overlapping lines,
lines within molecular bands and lines affected by more complex physical
effects can exhibit opposite or generally different behavior. As a result,
broad molecular bands visible even on a low resolution spectrum of a given
source may behave in different ways. The pattern of their changes depends
on physical parameters, composition and structure of the atmosphere.
Observations of these bands can therefore be used to constrain the global
parameters, and to check theoretical assumptions used for constructing
model stellar atmospheres. High resolution spectra resolving individual
lines would, however, also enable studies of the depth structure of the
star's atmosphere in detail. Particularly in the case of red giants,
gravitational microlensing therefore provides a unique new tool for
stellar physics.

During M95-30, the only well-documented point-mass microlensing transit
event up to now, eleven low resolution and three high resolution spectra
were taken in addition to photometry in two colors (Alcock et~al.~1997d).
In a forthcoming paper (Heyrovsk\'y \& Sasselov~2000b), we will use the
methods and findings described here together with the observational data
to study the lensed giant and construct an adequate model of its
atmosphere.

\acknowledgements 
We thank R. Kurucz for his kind help with opacities and molecular data. We
also thank an anonymous referee for comments that helped improve the
manuscript. DS is grateful to I. Shapiro for his support during part of
this study.


\begin{references}

\reference{}
Afonso, C., et al. 1999, A\&A, 344, L63
\reference{}
Albrow, M., et al. 1999, ApJ, 522, 1011
\reference{}
Alcock, C., et al. 1997a, ApJ, 479, 119
\reference{}
Alcock, C., et al. 1997b, ApJ, 486, 697
\reference{}
Alcock, C., et al. 1997c, ApJ, 491, L11
\reference{}
Alcock, C., et al. 1997d, ApJ, 491, 436
\reference{}
Andersen, J. 1991, A\&A Review, 3, 91
\reference{}
Armstrong, J. T., et al. 1995, Physics Today, 48, 42
\reference{}
Becker, A., et al. 1998, BAAS, 30, 1415
\reference{}
Bessell, M. S., Brett, J. M., Scholz, M., \& Wood, P. R. 1989, A\&AS, 77,
1 
\reference{}
Bessell, M. S., Brett, J. M., Scholz, M., \& Wood, P. R. 1991, A\&AS, 89,
335  
\reference{}
Bogdanov, M. B., \& Cherepashchuk, A. M. 1995, ARep, 39, 779
\reference{}
Bogdanov, M. B., \& Cherepashchuk, A. M. 1996, ARep, 40, 713
\reference{}
Bogdanov, M. B., Cherepashchuk, A. M., \& Sazhin, M. V. 1996, Ap\&SS, 235,
219
\reference{}
Gaudi, B. S., \& Gould, A. 1999, ApJ, 513, 619
\reference{}
Gould, A. 1994, ApJ, 421, L71
\reference{}
Gould, A. 1995, ApJ, 446, L71
\reference{}
Gould, A. 1997, ApJ, 483, 98
\reference{}
Gould, A., \& Welch, D. L. 1996, ApJ, 464, 212
\reference{}
Hauschildt, P. H., Allard, F., Alexander, D. R., \& Baron, E. 1997, ApJ, 
488, 428 
\reference{}
Hauschildt, P. H., Allard, F., Ferguson, J., Baron, E., \& Alexander, D.
R. 1999, ApJ, 525, 871 
\reference{}
Hendry, M. A., Coleman, I. J., Gray, N., Newsam, A. M., \& Simmons, J. F. 
L. 1998, NewAR, 42, 125
\reference{}
Heyrovsk\'y, D., \& Loeb, A. 1997, ApJ, 490, 38 (Paper I)
\reference{}
Heyrovsk\'y, D., \& Sasselov, D. D. 2000a, ApJ, in press
\reference{}
Heyrovsk\'y, D., \& Sasselov, D. D. 2000b, in preparation
\reference{}
Hofmann, K.-H., \& Scholz, M. 1998, A\&A, 335, 637
\reference{}
Houdashelt, M. L., Bell, R. A., Sweigart, A. V., \& Wing, R. F. 2000, AJ,
in press, preprint (astro-ph/9911383)
\reference{}
Ignace, R., \& Hendry, M. A. 1999, A\&A, 341, 201
\reference{}
Jacob, A. P., Bedding, T. R., Robertson, J. G., \& Scholz, M. 1999, MNRAS,
in press, preprint (astro-ph/9911076)
\reference{}
Kurucz, R. L. 1992, in IAU Symp. 149, The Stellar Populations of Galaxies,
ed. B. Barbuy \& A. Renzini (Dordrecht: Kluwer), 225 
\reference{}
Kurucz, R. L. 1999, in IAU Gen.Ass. XXIII, Highlights of Astronomy
Vol.11B, ed. J. Andersen (Dordrecht: Kluwer), 646
\reference{}
Lennon, D. J., Mao, S., Fuhrmann, K., \& Gehren, T. 1996, ApJ, 471, L23 
\reference{} 
Loeb, A., \& Sasselov, D. 1995, ApJ, 449, L33
\reference{}
Magain, P. 1986, A\&A, 163, 135
\reference{}
Maoz, D., \& Gould, A. 1994, ApJ, 425, L67
\reference{}
Mozurkewich, D., et al. 1991, AJ, 101, 2207
\reference{}
Nemiroff, R. J., \& Wickramasinghe, W.A.D.T. 1994, ApJ, 424, L21
\reference{}
Paczy\'nski, B. 1996, ARA\&A, 34, 419
\reference{}
Plez, B., Brett, J. M., \& Nordlund, A. 1992, A\&A, 256, 551
\reference{}
Renault, C., et al. 1997, A\&A, 324, L69
\reference{}
Sasselov, D. D. 1996, in ASP Conf. Ser. 109, Cool Stars 9, ed. R.
Pallavicini \& A. K. Dupree (San Francisco: ASP), 541
\reference{}
Scholz, M., \& Tsuji, T. 1984, A\&A, 130, 11
\reference{}
Schwenke, D. W. 1998, in Faraday Discussions 109, Chemistry and Physics of
Molecules and Grains in Space, ed. The Faraday Division of the Royal
Society of Chemistry (London: Royal Society of Chemistry), 321
\reference{}
Simmons, J. F. L., Newsam, A. M., \& Willis, J. P. 1995, MNRAS, 276, 182
\reference{}
Simmons, J. F. L., Willis, J. P., \& Newsam, A. M. 1995, A\&A, 293, L46
\reference{}
Udalski, A., et al. 1997, Acta Astron., 47, 169
\reference{}
Uitenbroek, H., Dupree, A. K., \& Gilliland, R. L. 1998, AJ, 116, 2501
\reference{}
Valls-Gabaud, D. 1995, in Large Scale Structure in the Universe, ed. J. P.
M\"ucket, S. Gottl\"ober, \& V. M\"uller (Singapore: World Scientific),
326
\reference{}
Valls-Gabaud, D. 1996, in IAU Symp. 173, Astrophysical Applications of
Gravitational Lensing, ed. C. S. Kochanek \& J. N. Hewitt (Dordrecht:
Kluwer), 237
\reference{}
Valls-Gabaud, D. 1998, MNRAS, 294, 747
\reference{}
Vandenberg, D., \& Bell, R. A. 1985, ApJS, 58, 561
\reference{}
Witt, H. J. 1995, ApJ, 449, 42
\reference{}
Witt, H. J., \& Mao, S. 1994, ApJ, 430, 505

\end{references}
\end{document}